\documentclass[iop]{emulateapj} 
\usepackage{natbib,amsmath,booktabs,apjfonts,threeparttable,multirow}
\newcommand{\snrs}{W41, G337.0--0.1, and MSH~17--3{\it 9}}
\newcommand{\msh}{MSH 17--3{\it 9}}
\newcommand{\wf}{W41}
\newcommand{\ctbt}{G337.0--0.1}
\newcommand{\magj}{ Swift J1834.9--0846}
\newcommand{\lef}{{\it left}}
\newcommand{\mi}{{\it middle}}
\newcommand{\rig}{{\it right}}

\newcommand{\xmm}{{\it XMM--Newton}}
\newcommand{\chandra}{{\it Chandra}}
\newcommand{\fermi}{{\it Fermi}-LAT}

\shortauthors{Castro et al.}
\shorttitle{FERMI-LAT OBSERVATIONS OF SNRS INTERACTING WITH MOLECULAR CLOUDS}
\slugcomment{Accepted for publication in ApJ}

\begin{document} 
\title{{\it Fermi} LAT Observations of Supernova Remnants Interacting \\with Molecular Clouds: W41, MSH~17--3{\it 9}, and G337.0--0.1}
\author{Daniel Castro\altaffilmark{1}, Patrick Slane\altaffilmark{2}, Ashley Carlton\altaffilmark{2}, Enectali Figueroa--Feliciano\altaffilmark{1}}

\altaffiltext{1}{MIT-Kavli Center for Astrophysics and Space Research, 77 Massachusetts Avenue, Cambridge, MA, 02139, USA}
\altaffiltext{2}{Harvard-Smithsonian Center for Astrophysics, 60 Garden Street, Cambridge, MA 02138, USA}

\begin{abstract}
We report the detection of $\gamma$-ray emission coincident with three supernova remnants (SNRs) using data from the Large Area Telescope on board the {\it Fermi Gamma-ray Space Telescope}. \snrs\ are SNRs known to be interacting with molecular clouds, as evidenced by observations of hydroxyl (OH) maser emission at 1720 MHz in their directions and other observational information. SNR shocks are expected to be sites of cosmic ray acceleration, and clouds of dense material can provide effective targets for production of $\gamma$-rays from $\pi^0$-decay. The observations reveal unresolved sources in the direction of G337.0--0.1, and MSH~17--3{\it 9}, and an extended source coincident with \wf. We model their broadband emission (radio to $\gamma$-ray) using a simple one-zone model, and after considering scenarios in which the MeV-TeV sources originate from either $\pi^0$-decay or leptonic emission, we conclude that the $\gamma$-rays are most likely produced through the hadronic channel.

\end{abstract}

\keywords{acceleration of particles --- cosmic rays --- gamma rays: ISM  --- ISM: individual (W41, MSH~17--3{\it 9}, G337.0--0.1) --- ISM: supernova remnants}

\section{Introduction}

\noindent One hundred years after being detected for the first time, the detailed nature of Galactic cosmic ray (CR) acceleration is yet to be definitely understood. The observational evidence supporting diffusive shock acceleration (DSA) at SNR shocks as the primary origin of Galactic CRs is reviewed in \citet{reynolds_2008}, and includes the detection of non-thermal X-ray emission from young shell-type SNRs, such as RX J1713.7-3946 \citep{koyama_1997,slane_1999}, and Vela Jr. \citep{aschenbach_1998,slane_2001}. These X-rays are believed to be synchrotron radiation from electrons accelerated to TeV energies at the SNR shock. Observations of $\gamma$-ray emission in the very high energy (VHE) $\gamma$-ray range from some SNRs also support the scenario where particles are being accelerated to energies approaching 10$^{15}$ eV in these objects \citep{muraishi_2000,aharonian_2004,katagiri_2005}. It has proven difficult to determine, however,  whether these high energy photons result from leptonic emission mechanisms (inverse Compton scattering or non-thermal bremsstrahlung emission), or from relativistic hadrons interacting with the ambient medium \citep[e.g.][]{ellison_2010,inoue_2012}. 

Studies of the hydrodynamic evolution of SNRs and the properties of the shocked medium and ejecta have also helped characterize CR production. The X-ray morphologies of Tycho's SNR and SN~1006 indicate that the shock compression ratios in these objects are higher than expected due to efficient particle acceleration at their shocks \citep{warren_2005,cassam_2008}. Postshock plasma temperatures estimated from observations of SNRs 1E~0102.2-7219 and RCW~86 are cooler than those expected from the shock velocities measured, also suggesting that a significant fraction of their explosion energy has been placed in relativistic particles \citep{hughes_2000,helder_2009}. 

The interaction of CRs with regions of high-density material is expected to result in $\gamma$-ray emission from the decay of neutral pions generated by proton-proton interactions, as suggested in \citet{claussen_1997}. The spectral energy distribution from this emission process is characterized by sharply rising in the $\sim 70-200$ MeV range, and tracing the parent accelerated ion population above $\sim1$ GeV \citep{kamae_2006}. Hence, when searching for signatures of CR acceleration in remnants, those with shocks believed to be propagating into molecular clouds (MC) make an excellent laboratory. Recent observations with the {\it Fermi} Large Area Telescope, as well as AGILE/GRID, have shown that SNR--MC systems are an important class of MeV-GeV $\gamma$-ray sources, also suggested by correlation studies carried out by \citet{hewitt_2009}. Since its launch in 2008, \fermi\ has allowed successful detections of several such systems in the MeV-GeV energy range, such as SNRs W51C \citep{abdo_w51c}, G349.7--0.5, CTB~37A, 3C~391, and G8.7--0.1 \citep{castro_snrmc}. Recently, observations of the SNR W44 with AGILE/GRID combined with Fermi-LAT data led \citet{giuliani_2011} to conclude that the characteristics of the resulting spectrum indicate the $\gamma$-ray emission originates from $\pi^0$-decay of CR hadrons interacting with ambient material. Moreover, studies of SNRs IC443 and W44 by \citet{ackermann_2013} extended the spectral information gathered with \fermi\ down to 60 MeV, allowing the  $\pi^0$-decay feature to be clearly traced, and thus identifying the $\gamma$-ray emission as predominantly hadronic in origin. In contrast, observations of SNRs RX J1713.7-3946 and Vela Jr. with \fermi\ together with broadband data have been used to argue that the $\gamma$-rays detected in their direction are produced through leptonic processes \citep{abdo_rxj1713,ellison_2010,lee_2013}.

SNRs \snrs\ show evidence of shock interaction with dense local molecular clouds. In this paper we  study the $\gamma$-ray emission in the direction of these SNRs, and establish constraints on the nature of these systems through modeling of their broadband spectra. We report the detection of three \fermi\ sources coincident with these supernova remnants, and on the results of our detailed study of $\gamma$-ray emission in their regions. 
 
\section{OBSERVATIONS AND DATA ANALYSIS}

\noindent Data from 52 months of observations with the {\it Fermi Gamma-ray Space Telescope} Large Area Telescope ({\it Fermi}-LAT) (from 4 August, 2008 until 30 November 2012) were obtained and  studied in this work. Exclusively, events belonging to the Pass 7 V6 {\it Source} class, selected to reduce the residual background rate as explained in detail in \citet{abdo_p7}, have been used in the analysis presented. The updated instrument response functions (IRFs) used are ``Pass7 version 6'', obtained using data gathered since the launch of the observatory \citep{rando_2009,abdo_p7}. Only events from zenith angles $\phi>100^{\circ}$ are selected so as to reduce the terrestrial albedo $\gamma$-ray flux \citep{abdo_prd}. The analyses of each of the SNRs include data from circular regions of the sky centered on each object, with radius 25$^\circ$.

The  $\gamma$-ray data in the direction of the SNRs of interest are analyzed using the Fermi Science Tools v9r27p1\footnote{The Science Tools package and support documents are distributed by the Fermi Science Support Center and can be accessed at http://fermi.gsfc.nasa.gov/ssc}. We employ the maximum likelihood fitting technique, and use {\it gtlike} to constrain the morphological and spectral characteristics of each region \citep{mattox_1996}. The source models incorporated in the {\it gtlike} analyses include a Galactic diffuse component, which results from the interactions of the field of cosmic rays with interstellar material and photons, and an isotropic component that accounts for the extragalactic diffuse and residual $\gamma$-ray backgrounds. To this effect we use the mapcube file {\tt gal\_2yearp7v6\_v0.fits} to describe the $\gamma$-ray emission from the Milky Way, and model the isotropic component using table {\tt iso\_p7v6source.txt}, both of which can be obtained from the Fermi Science Support Center. 

To understand the spatial properties of the $\gamma$-ray radiation in the field of \snrs, we use \fermi\ data in the energy range 2 to 200 GeV, and select only events detected in the {\it front} section of the instrument. This allows for a good compromise between photon numbers and spatial resolution, since the 68\% containment radius angle for normal incidence {\it front}-selected photons in this energy band is $\leq 0.3^\circ$, and much larger for photons with lower energies. Galactic and isotropic backgrounds are modeled and test statistic (TS) maps are constructed through {\it gttsmap}, and hence the detection significance is estimated for each of the objects. This is also useful for evaluating the position and possible extent of each source.  The test statistic is the logarithmic ratio of the likelihood of a point source being at a given position in the sky, to the likelihood of the model without that additional source, $2\text{log}(L_{\text{ps}}/L_{\text{null}})$.

The $\gamma$-ray spectral energy distribution (SED) characteristics of the sources coincident with each supernova remnant studied are determined using events converted in both {\it front} and {\it back} sections, and for photons within the energy range 0.2-204.8 GeV. The lower energy limit is chosen because the effective area of the instrument at low energies changes very rapidly with energy, and because of the large uncertainty in the Galactic diffuse background below 200 MeV. {\it gtlike} is used to model the $\gamma$-ray flux in each of the 10 logarithmically spaced energy bins and estimate, through the maximum likelihood technique, the best-fit parameters for all the sources modeled. Background sources from the 24-month {\it Fermi} LAT Second Source Catalog \citep{nolan_2012}\footnote{ The data for the 1873 sources in the {\it Fermi} LAT Second Source Catalog is made available by the Fermi Science Support Center at http://fermi.gsfc.nasa.gov/ssc/data/access/lat/2yr\_catalog/} have been considered in the model likelihood fits. For the IRFs used in our study, the systematic uncertainties of the effective area, vary with energy: 10\% at 100 MeV, decreasing to 5\% at 560 MeV, and increasing to 20\% at 10 GeV \citep[and references therein]{porter_2009,nolan_2012}. In addition to the statistical uncertainties associated with the likelihood fits to the data, and the systematic errors related to the IRFs, the uncertainty of the Galactic diffuse background intensity is considered. To take this into account in the total systematic uncertainties we vary the normalization of the Galactic background by $\pm6\%$ from the best-fit values at each of the energy bins and estimate the flux from the object of interest using these new artificially frozen values of the background, as outlined in \citep{castro_snrmc}.

\subsection{W41}

\noindent The supernova remnant W41 (G23.3-0.3) is observed as an asymmetric partial radio shell, nearly overlapping SNR G22.7-0.2, which is located to its south \citep{kassim_1992}. The Very Large Array Galactic Plane Survey (VGPS) observations at 1420 MHz reveal an angular size for W41 of 36$'\times30'$, and combined with the 330 MHz study from \citet{kassim_1992}, yield a spectral index lower limit of --0.43 \citep{tian_2007}. \citet{green_1997} report the detection of emission at 1720 MHz from this SNR, yet it was not confirmed as maser emission with their subsequent high resolution interferometer studies. \citet{albert_2006} observed a giant molecular cloud (GMC) coincident with this region, by studying the $^{12}$CO data. Emission spectra from $^{13}$CO shows that a GMC, at velocity 77 km s$^{-1}$, is detected over the projected extent of the SNR.

\begin{figure*}
\begin{center}
\includegraphics[width=0.3\textwidth]{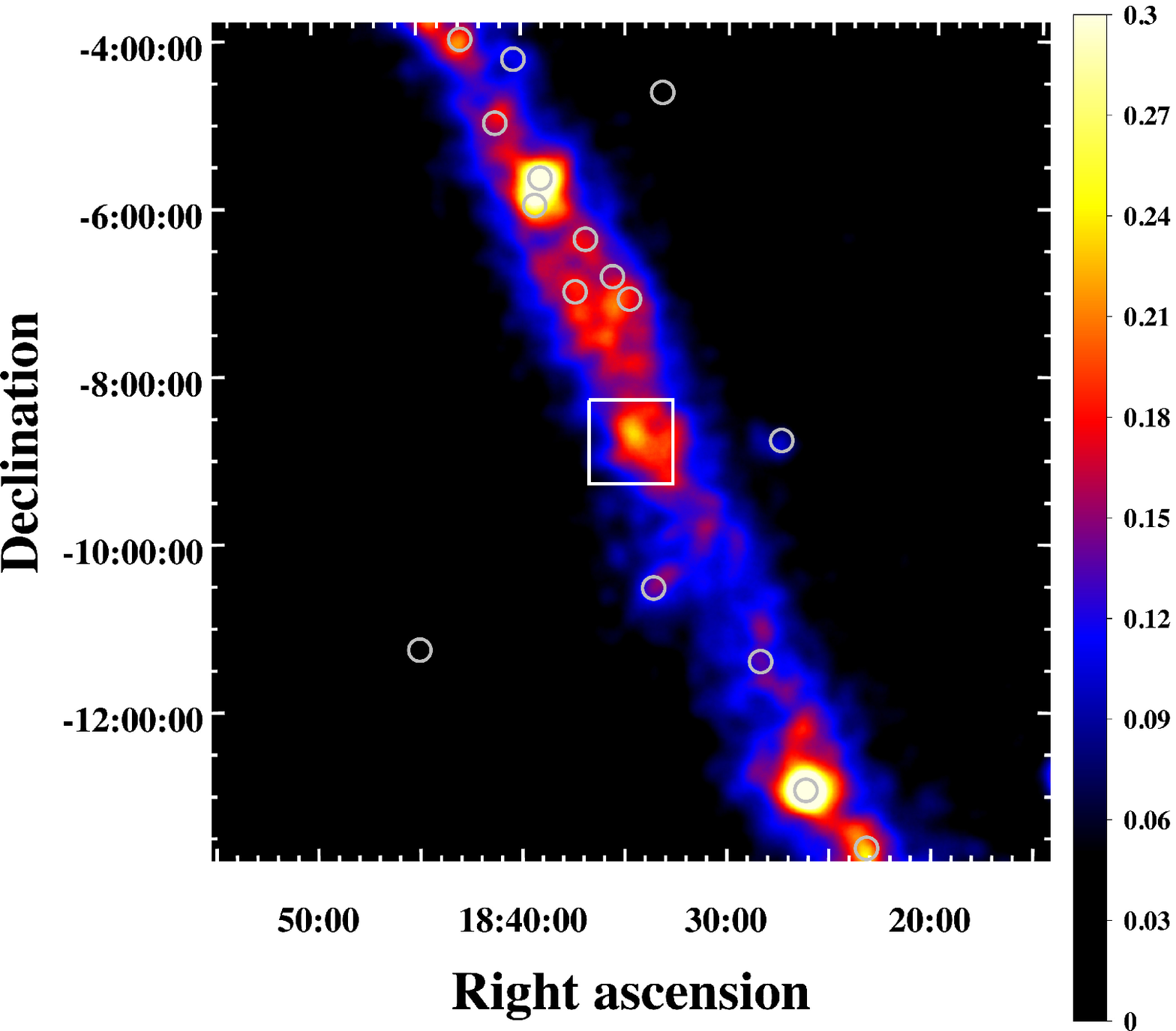}
\includegraphics[width=0.3\textwidth]{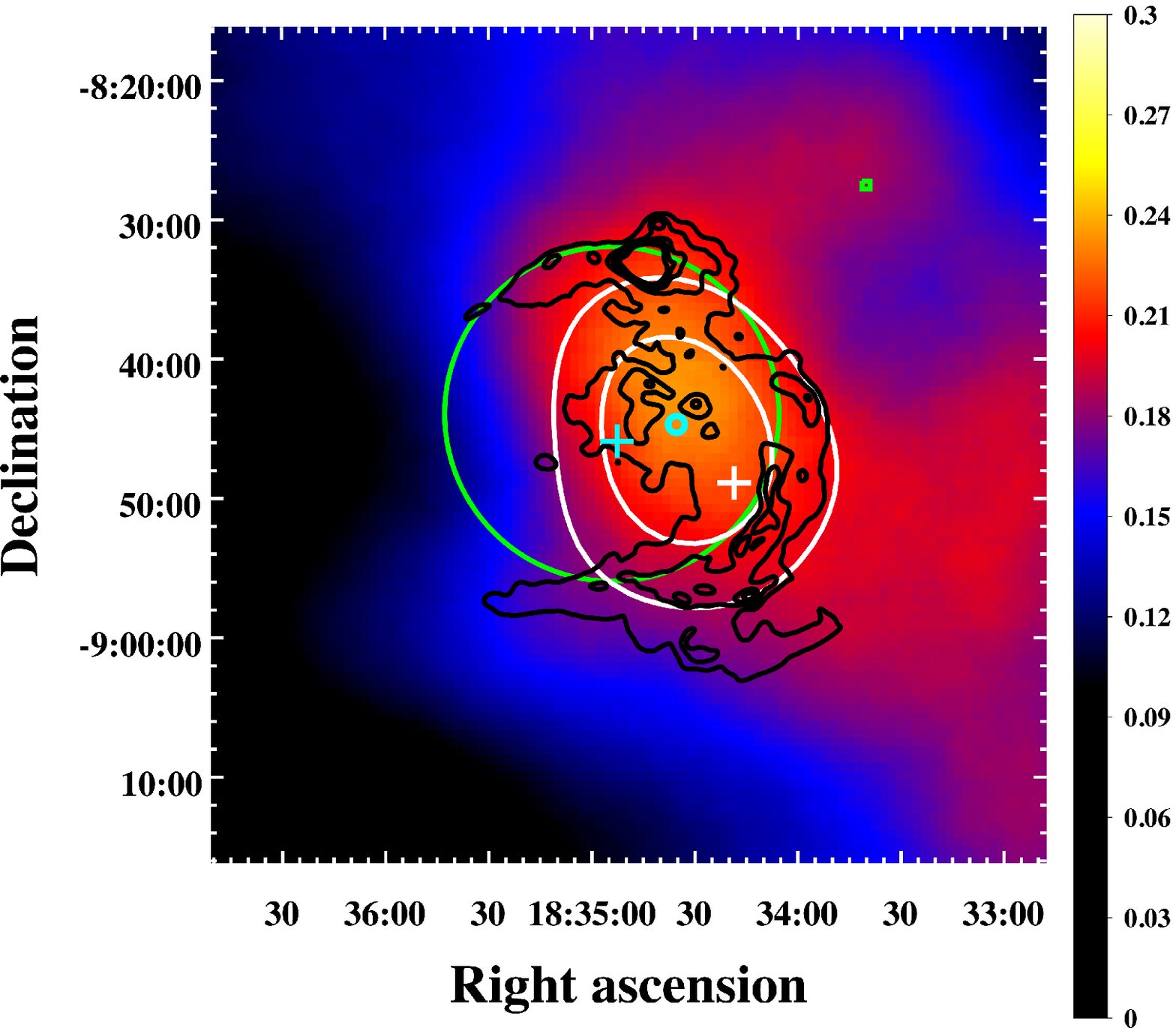}
\includegraphics[width=0.35\textwidth]{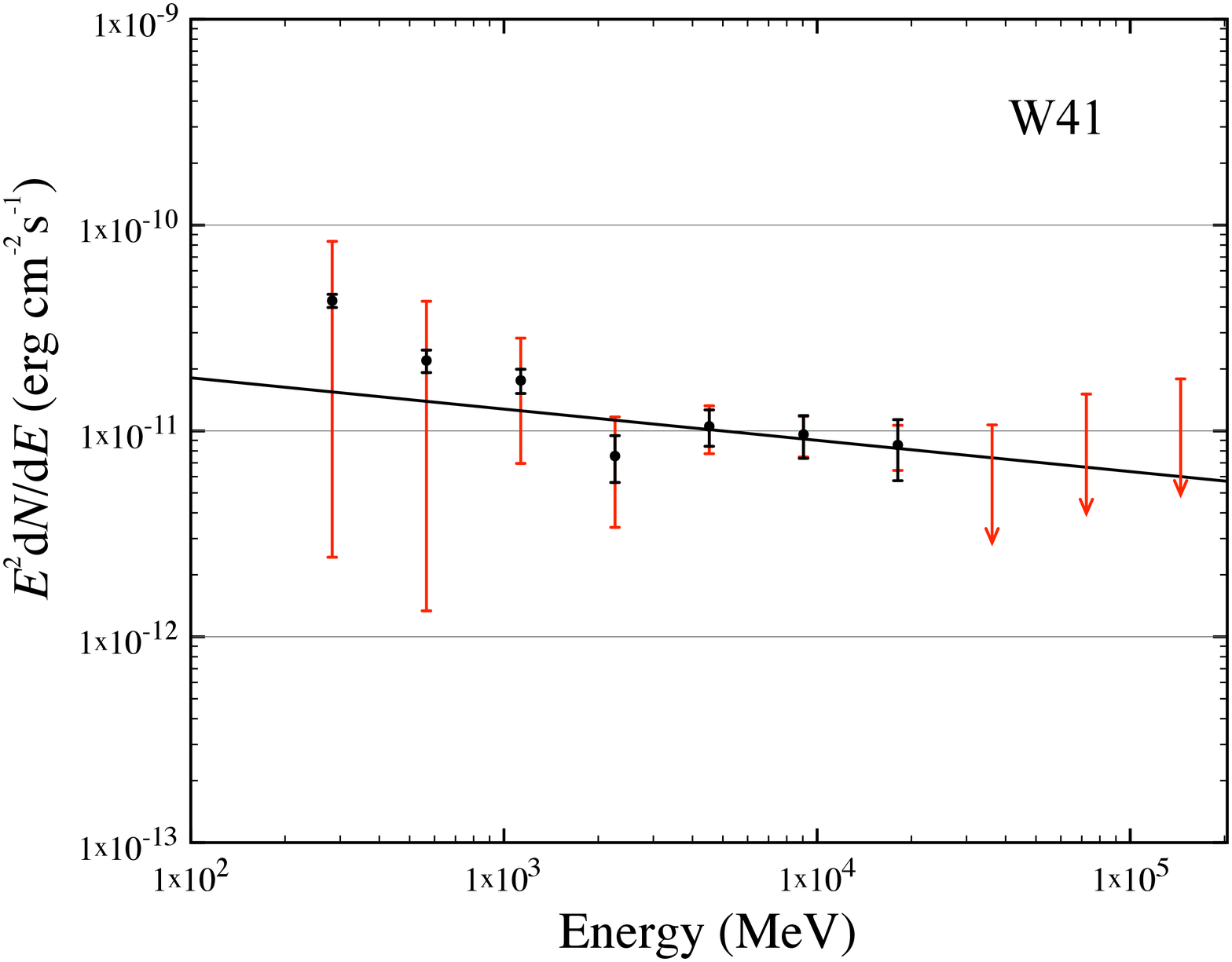}\\
\caption{  \footnotesize The {\it left} and {\it middle} panels show count maps of {\it front} converted events in the range 2 to 200 GeV around SNR \wf. Units are $10^4 \text{counts deg}^{-2}$, the pixel binning is 0.01$^\circ$, and the maps are smoothed through convolution with a Gaussian distribution of width 0.2$^\circ$. The {\it left} panel shows a $10^\circ\times 10^\circ$ region, where grey circles indicate the positions of 2FGL catalogued sources in the field \citep{nolan_2012}, and the white square shows the $1^\circ\times 1^\circ$ region shown in the {\it middle} panel.
The black contours show the radio emission, from VGPS observations at 1420 MHz. The white cross represents the position of the catalogued $\gamma$-ray source 2FGL~J1834.3--0848 \citep{nolan_2012}, and the green circle shows the approximate TeV $\gamma$-ray extension of HESS J1834--087. The green box shows the position of the old pulsar PSR J1833-087 \citep{gaensler_1995}, the cyan circle encloses the position of the hard X-ray sources detected within the central region of W41 \citep{mukherjee_2009,misanovic_2011}, and the cyan cross marks the position of the magnetar \magj \citep{kargaltsev_2012,younes_2012}. The white curves show the TS$=49$ and TS$=64$ contours (which correspond to significance 7$\sigma$ and 8$\sigma$). The  {\it right} panel shows the {\it Fermi} LAT spectral energy distribution of the source coincident with SNR \wf. Statistical uncertainties are shown as black error bars, and systematic errors are indicated by red bars. The solid line represent the best-fit powerlaw model to the data.}
\label{fig:w41}
\end{center}
\end{figure*}

Observations in the TeV range with HESS have revealed a bright source, HESS J1834--087, 12$'$ in diameter, positionally coincident with W41 \citep{aharonian_2006}. Additionally, {\it XMM-Newton} observations reveal diffuse X-ray emission coincident with the HESS source \citep{tian_2007}. The coincidence between the CO data, the X-ray emission and the TeV source suggest that the interaction between the SNR shock and the GMC, where the shock-accelerated hadrons interact with the dense material and undergo $\pi^0$-decay, is the origin of the $\gamma$-ray emission. Using this association, \citet{leahy_2008} establish a kinematic distance to W41 of $d=3.9-4.5$ kpc, using $^{13}$CO observations of the GMC. Subsequent studies with \xmm, \chandra\ and {\it Swift} have detected X-ray point sources coincident with HESS J1834--087 \citep{mukherjee_2009,misanovic_2011}, including a magnetar, Swift J1834.9--0846 \citep{delia_2011}. Additionally, there is a radio-detected pulsar located $\sim20'$ NW of the center of W41 \citep{gaensler_1995}. We discuss the nature of these sources and possible relation to the \fermi\ and HESS detected emission in $\S$3.2.2.

The smoothed count map from the \fermi\ data, in the region surrounding W41, is shown in Figure \ref{fig:w41} (\lef\ and \mi\ panels). The TS contours clearly show that the emission is coincident with the radio extent of the remnant. The {\it Fermi} LAT source 2FGL~J1834.3--0848 is also coincident with \wf, and its position is shown in Figure \ref{fig:w41} (\mi\ panel).  This source is reported to have an estimated power law index $\Gamma=2.4\pm 0.2$ and flux $7.9\pm0.7\times10^{-11} $ erg cm$^{-2}$ s$^{-1}$, in the 0.1 to 100 GeV energy range \citep{nolan_2012}. 
 
\begin{figure}[h!]
\begin{center}
\includegraphics[width=\columnwidth]{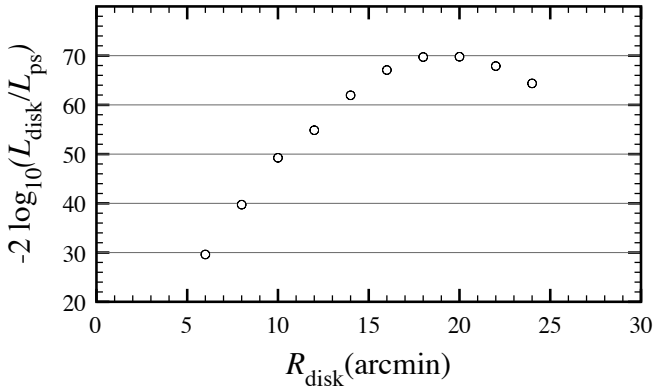}
\caption{  \footnotesize Logarithmic ratio between the likelihood of disk-like spatial distributions of emission in the direction of \wf\ and the likelihood of a point-source description (test statistic of a disk template), as a function of disk radius. The distribution peaks at 20$'$, and its TS value suggests the $\gamma$-ray source is significantly extended.}
\label{fig:profile}
\end{center}
\end{figure}

The possible extension of this source was further explored by comparing the overall likelihood obtained with {\it gtlike} for spatial models of the emission from W41 as disk templates spanning several sizes ($r_{\text{disk}}=6'-12'-18'-24'$) with the likelihood of the point source model. 
As illustrated in figure \ref{fig:profile}, where the test-statistic values for each disk size are shown, the extended templates are more successful at fitting the observations than the point source model, with TS peaking at $\sim 70$ for the $20'$ disk. Hence, we conclude that while we cannot rule out a point-like emission model, the morphology of the $\gamma$-ray source is possibly as extended as the SNR. 

The spectrum of the $\gamma$-ray emission is shown in Figure \ref{fig:w41} (\rig\ panel), where the best-fit power law model to the data is also presented. For energies above 25.6 GeV only flux upper limits are determined from the data. The best-fit model to the photon distribution yields a spectral index of $\Gamma = Ð2.2 \pm 0.2$, and an integrated flux between 100 MeV and 100 GeV of $F_{\gamma}\approx 9.8\times10^{-8}(7.7\times10^{-11})$ photons/cm$^{2}$/s (erg/cm$^{2}$/s). Using the kinematic distance estimated, 4.2 kpc, the luminosity in this bandpass is $L_{\gamma}\approx2\times10^{35}$ erg/s.

\subsection{\ctbt}

\noindent The small diameter SNR G337.0--0.1, together with several H{\footnotesize \,II} regions, forms the CTB 33 complex. ATCA continuum observations at 1380 MHz reveal that G337.0--0.1 has a non-thermal spectrum, with spectral index -0.6, and shell-like morphology  \citep{sarma_1997}. Observations at 8.9 GHz by these authors reveal no (H90$\alpha$) emission, the recombination line produced by the transition between the n=91 and n=90 levels of hydrogen atoms and usually associated with thermal radio emission sources such as H~{\footnotesize II} regions. \citet{frail_1996} detected three OH(1720 MHz) maser spots in the direction of the CTB 33 complex, all with velocities $\sim 70$ km s$^{-1}$, one of which is directly coincident with G337.0--0.1. The non-thermal radio spectrum, combined with the recombination and maser information strongly points to the conclusion that G337.0--0.1 is a supernova remnant. The velocity of the maser emission, together with recombination line information from H{\footnotesize \,II} region G337.1-0.2, which is believed to be associated with \ctbt, place the SNR-H{\footnotesize \,II} region system at a distance of 11 kpc \citep{sarma_1997}.

\begin{figure*}
\begin{center}
\includegraphics[width=0.3\textwidth]{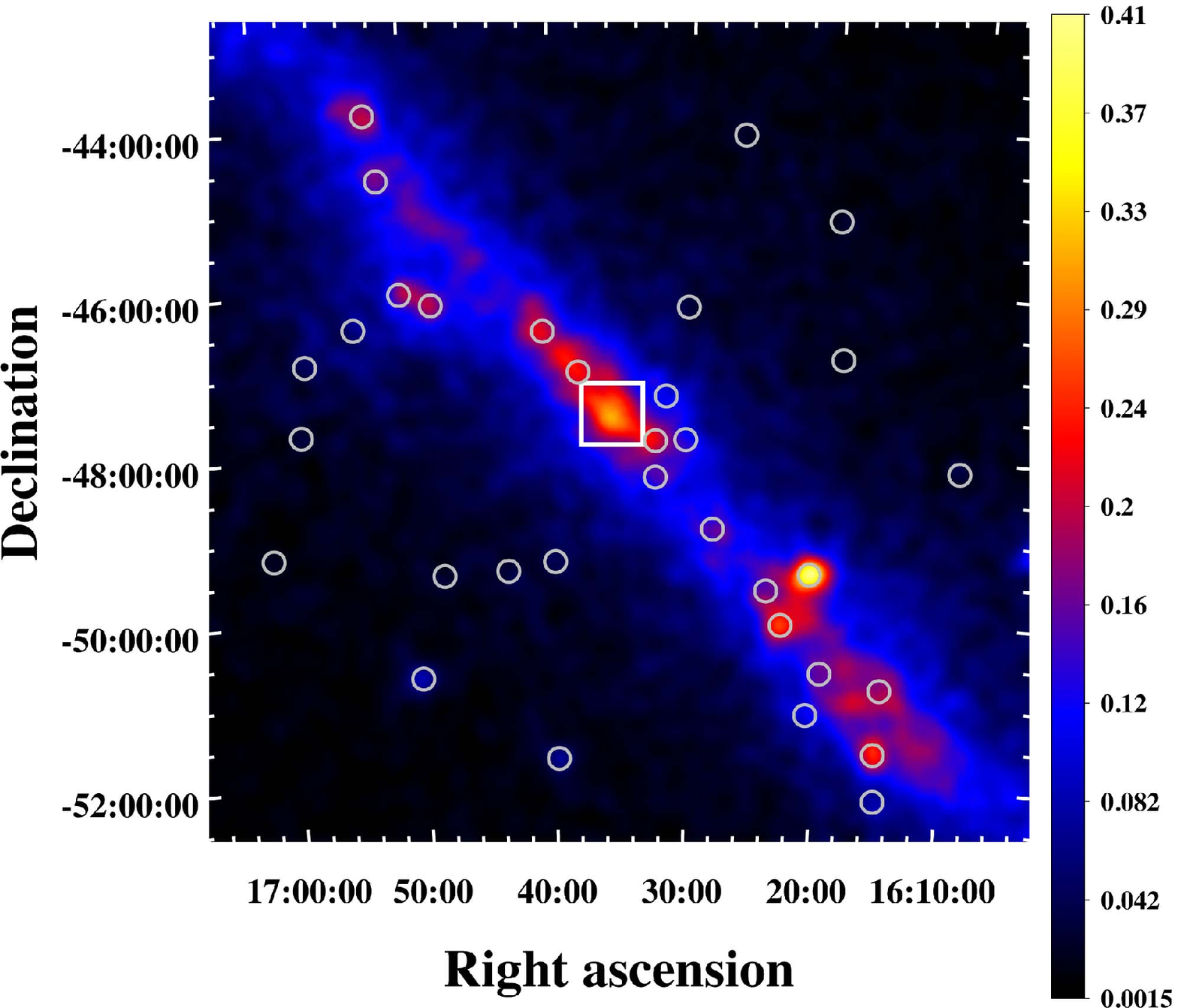}
\includegraphics[width=0.3\textwidth]{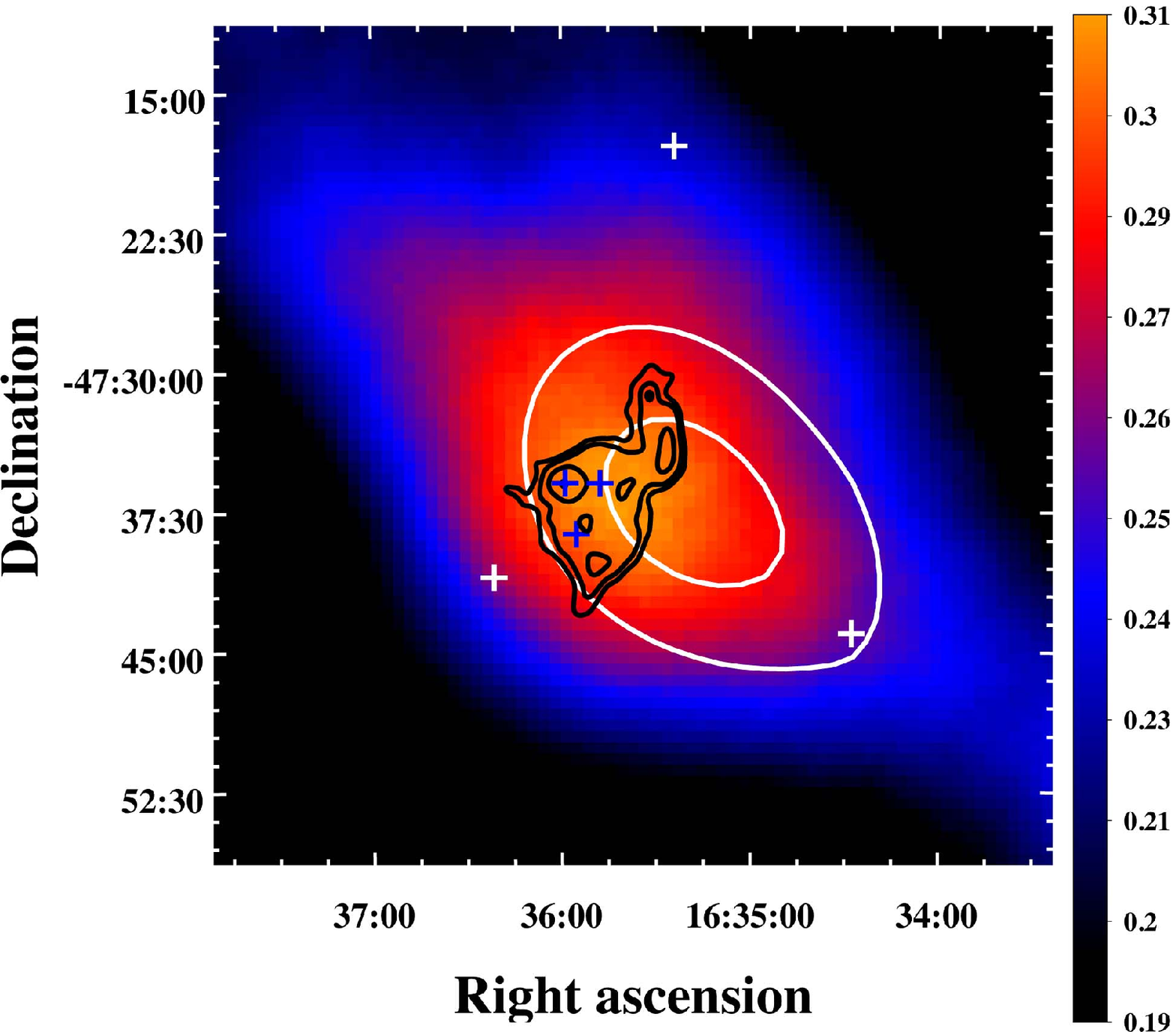}
\includegraphics[width=0.35\textwidth]{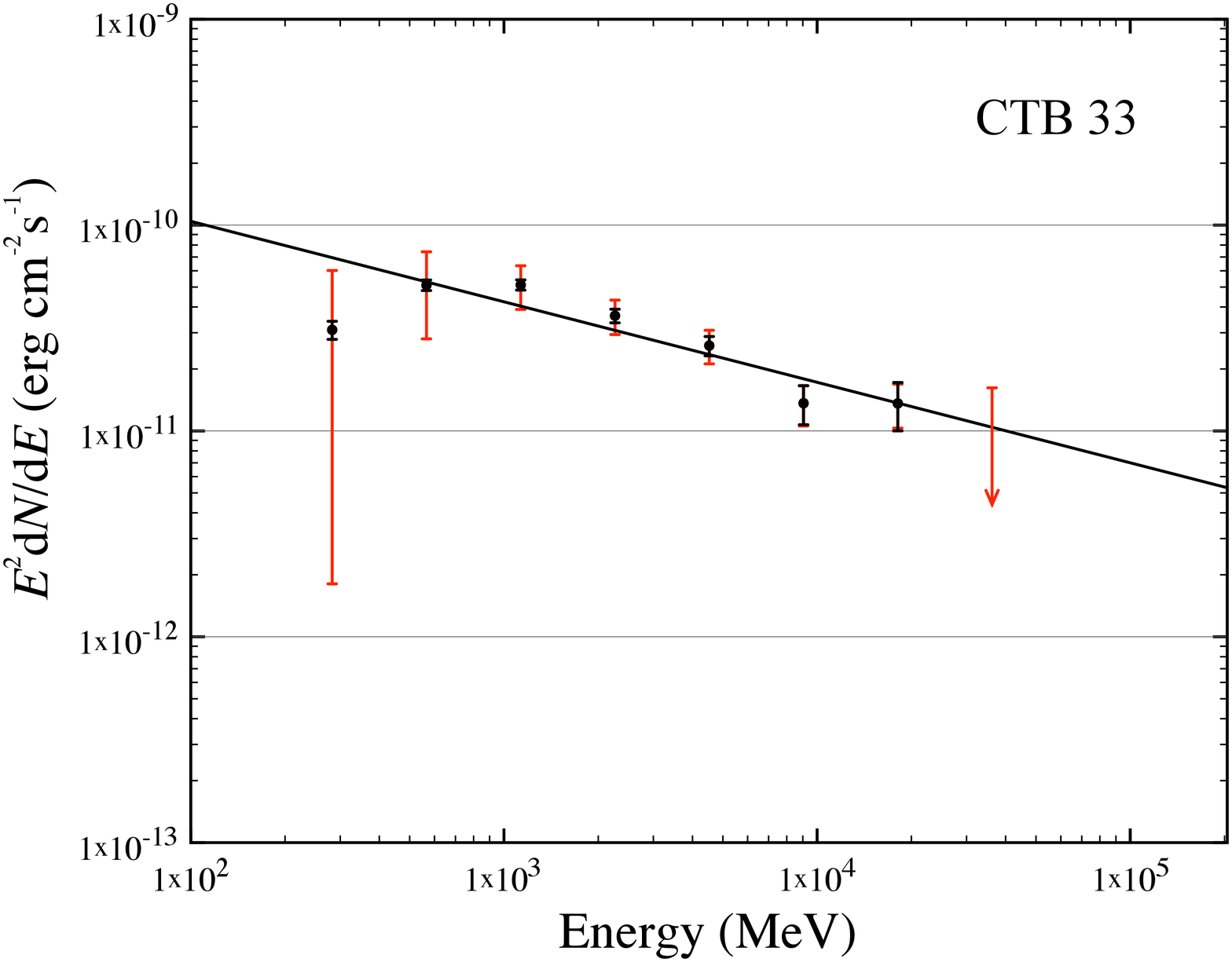}\\
\caption{  \footnotesize The {\it left} and {\it middle} panels show count maps around SNR \ctbt. Units are $10^4 \text{counts deg}^{-2}$, the pixel binning is 0.01$^\circ$, and the maps are smoothed through convolution with a Gaussian distribution of width 0.2$^\circ$. The {\it left} panel shows a $10^\circ\times 10^\circ$ region, where grey circles indicate the positions of 2FGL catalogued sources in the field\citep{nolan_2012}, and the white square shows the $45'\times 45'$ region shown in the {\it middle} panel. The black contours show the radio emission, from observations with MOST at 843 MHz. The white crosses represent the position of 2FGL sources in the field \citep{nolan_2012}, and the blue crosses indicate the positions of OH masers \citep{frail_1996}. The white curves show the test statistic 400 and 421 contours (which correspond to significance 20$\sigma$ and 21$\sigma$). The  {\it right} panel shows the {\it Fermi} LAT spectral energy distribution of the source coincident with SNR \ctbt. Statistical uncertainties are shown as black error bars, and systematic errors are indicated by red bars. The solid line represent the best-fit powerlaw model to the data.}
\label{fig:ctb33}
\end{center}
\end{figure*}

 \begin{figure*}
\begin{center}
\includegraphics[width=0.3\textwidth]{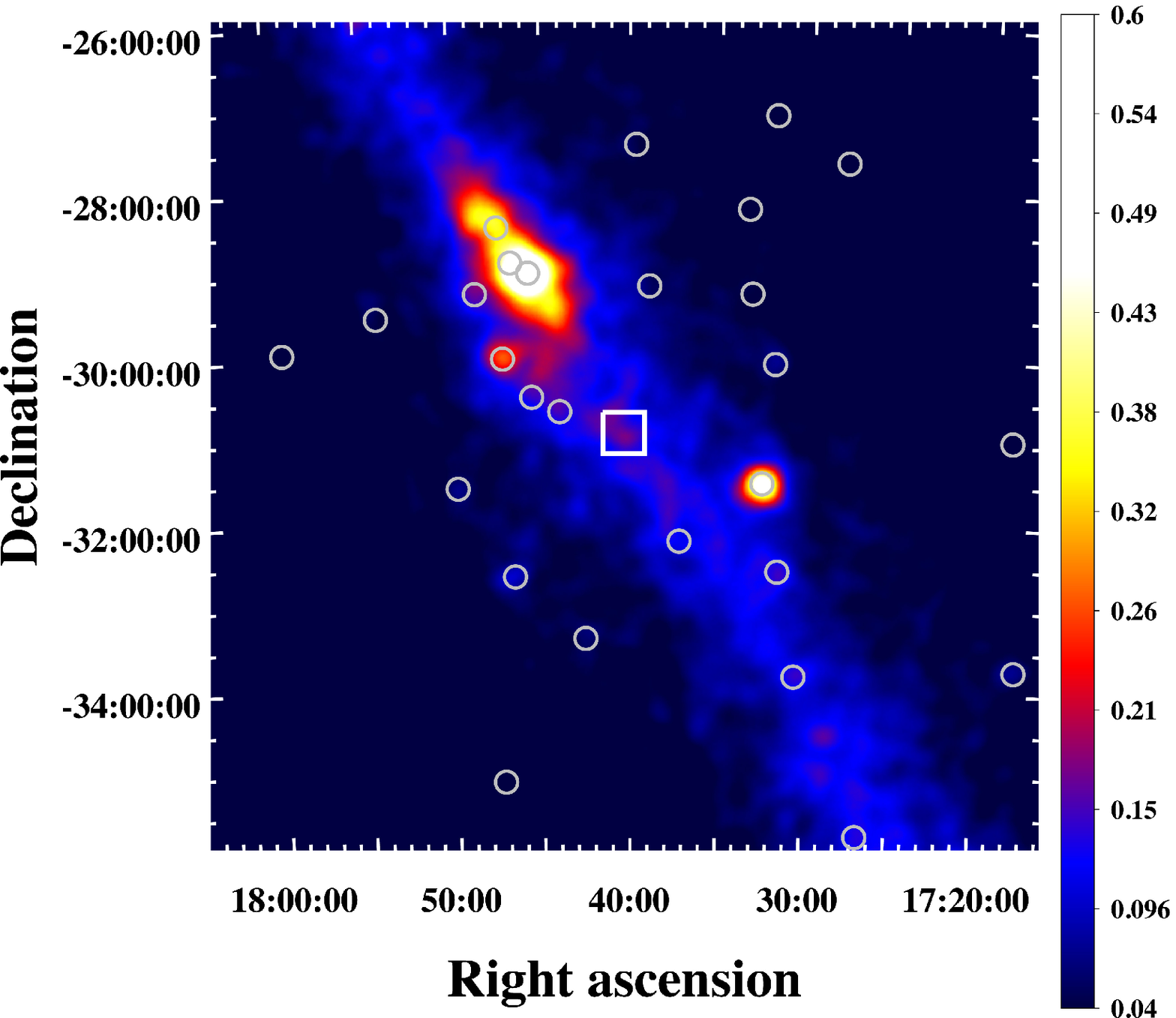}
\includegraphics[width=0.3\textwidth]{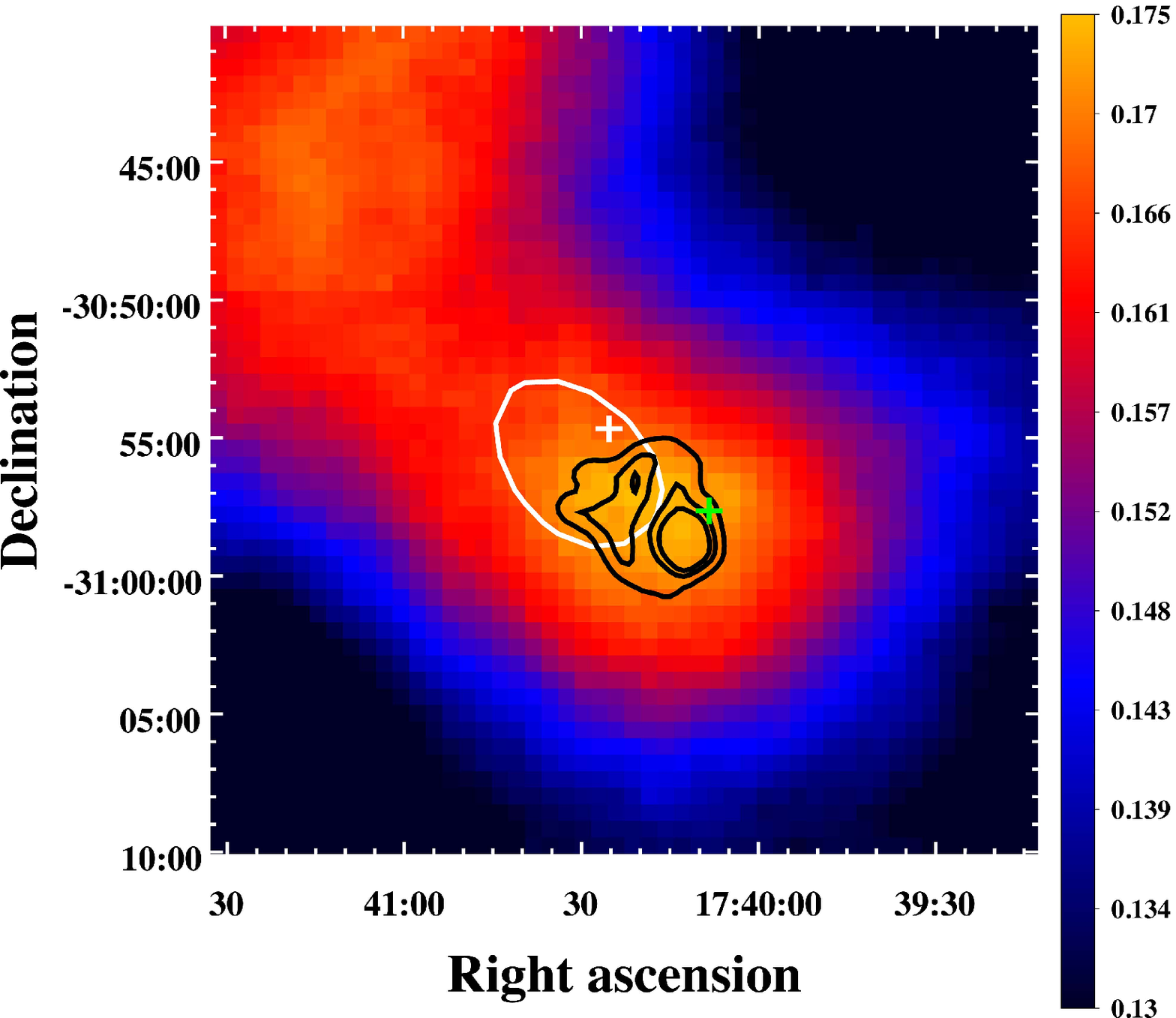}
\includegraphics[width=0.35\textwidth]{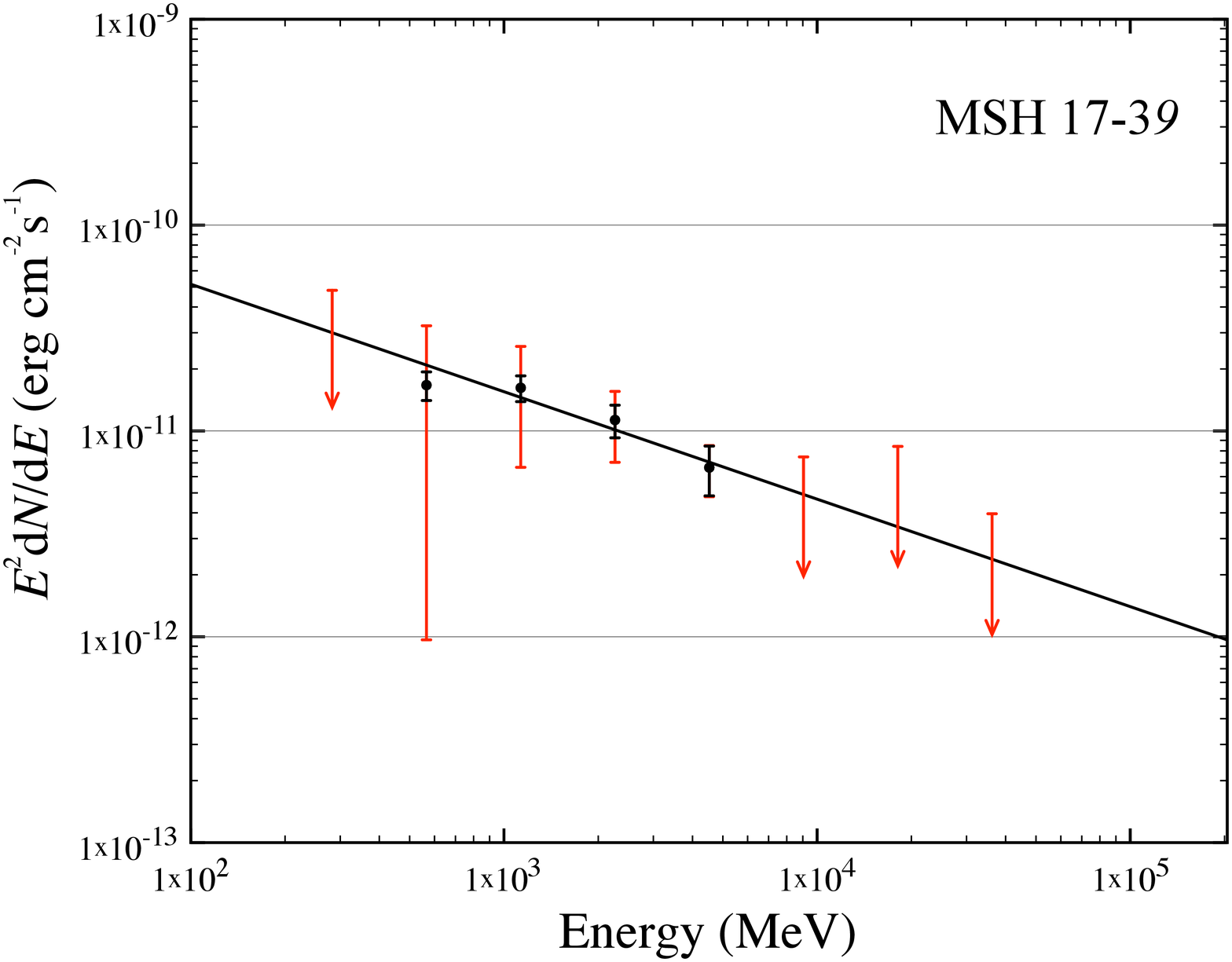}\\
\caption{  \footnotesize The {\it left} and {\it middle} panels show count maps of {\it front} converted events in the range 2 to 200 GeV around SNR \msh. Units are $10^4 \text{counts deg}^{-2}$, the pixel binning is 0.01$^\circ$, and the maps are smoothed through convolution with a Gaussian distribution of width 0.2$^\circ$. The {\it left} panel shows a $10^\circ\times 10^\circ$ region, where grey circles indicate the positions of 2FGL catalogued sources in the field \citep{nolan_2012}, and the white square shows the $30'\times 30'$ region shown in the {\it middle} panel.
The black contours show the radio emission from NVSS observations at 1420 MHz \citep{condon_1998}. The white cross represent the position of the catalogued $\gamma$-ray source 2FGL~J1740.4--3054c \citep{nolan_2012}, and the green cross indicates the positions of an OH maser \citep{frail_1996}. The white curve shows the test statistic 56.25 contour (which corresponds to significance 7.5$\sigma$). The  {\it right} panel shows the {\it Fermi} LAT spectral energy distribution of the source coincident with SNR \msh. Statistical uncertainties are shown as black error bars, and systematic errors are indicated by red bars. The solid line represent the best-fit powerlaw model to the data.}
\label{fig:msh1739}
\end{center}
\end{figure*}

\citet{woods_1999} report the observation with BATSE of a soft gamma repeater (SGR), SGR 1627-41, and its rough location suggests an association with SNR G337.0--0.1. The follow up X-ray observations of the region by these authors with $\it BeppoSAX$ also reveled the source SAX J1635.8-4736 consistent with the SNR position.

Figure \ref{fig:ctb33} (\lef\ and \mi\ panels) display {\it Fermi} LAT smoothed count maps of the region surrounding the CTB 33 complex, in the 2-200 GeV energy range. The overlaid TS contours suggest the $\gamma$-ray emission in the region is associated with G337.0--0.1. Additionally, the {\it Fermi} LAT First Source Catalog includes the source 2FGL J1636.4--4740, located at close proximity to G337.0-0.1. This 2FGL source is estimated to have distribution index $\Gamma=2.4\pm 0.1$ and flux $1.1\pm0.1\times10^{-10} $ erg cm$^{-2}$ s$^{-1}$, in the 0.1 to 100 GeV energy range \citep{nolan_2012}. 

The spectrum of the $\gamma$-ray emission is shown in Figure \ref{fig:ctb33} (\rig\ panel), where the best-fit power law model to the data is also presented. For energies above 25.6 GeV only flux upper limits are determined from the data. The best-fit model to the photon distribution yields a spectral index of $\Gamma = Ð2.4 \pm 0.1$, and an integrated flux between 100 MeV and 100 GeV of $F_{\gamma}\approx 4.7\times10^{-7}(2.5\times10^{-10})$ photons/cm$^{2}$/s (erg/cm$^{2}$/s). At the distance estimated from the maser velocity and recombination line observations, 11 kpc, the luminosity in this bandpass is hence estimated to be $L_{\gamma}\approx4\times10^{36}$ erg/s.

\subsection{\msh}

\noindent \msh\  (G357.7--0.1, Tornado Nebula) is a bright radio source classified as an SNR due to its steep non-thermal spectrum, high level of polarization and extended structure \citep[and references therein]{shaver_1985}. However, the axisymmetric morphology of this object, revealed in high resolution radio observations, is very peculiar and has led to variety of interpretations about its nature, including an extragalactic ``head-tail" source \citep{caswell_1989}, a synchrotron nebula powered by en energetic pulsar wind \citep{shull_1989}, and an accretion-powered nebula \citep{becker_1985}. The scenario where the peculiar structure of this object is a consequence of the SNR shock interacting with the surrounding dense molecular material finds support in the detection of a 1720 MHz OH maser, with velocity $-12.4$ km s$^{-1}$, coincident with the ``head" of the Tornado \citep{frail_1996}. \citet{lazendic_2003} provide further evidence of shock-molecular cloud interaction with observations of $^{13}$CO 1-0 and shock excited H$_2$ at 2.12$\mu$m at the position and velocity. H$_{\text{I}}$ absorption measurements \citep{brogan_2003} are consistent with the distance derived from the observed maser velocity, 12 kpc.

Observations with Chandra show three extended sources of X-ray emission in the direction of \msh, the brightest being located at the same position as that with the highest radio flux, the ``head" of the Tornado \citep{gaensler_2003}. The spectrum of this region in the X-ray band is highly absorbed (N$_{\text{H}}\approx 10^{23}\text{cm}^{-2}$) and is consistent with either a  collisionally ionized plasma, with temperature $\sim0.6$ keV, or a power law model ($\Gamma >4.5$).  The power law index established for the non-thermal emission model is rather steep and hence not consistent with the pulsar wind nebula (PWN) scenario \citep{gotthelf_2003}. 

The {\it Fermi} LAT $\gamma$-ray smoothed count map (in the 2-200 GeV band) of the region surrounding \msh\ is shown in Figure \ref{fig:msh1739}(\lef\ and \mi\ panels). Overlaid are the test statistic contours (TS$=-2\text{ln(}-L)$, where $L$ is the likelihood of the assumption of a point source being located at the position of the pixel), which clearly show a peak coincident with the position of the radio emission from \msh, and the maser spot. Also shown is the best-fit position of  source 2FGL~J1740.4--3054c,  which is located less than 0.1$^\circ$ North from the position of the ``head" of the Tornado. \citet{nolan_2012} estimate that this source has a power law index $\Gamma=2.4\pm 0.2$ and flux $7.0\pm0.8\times10^{-11} $ erg cm$^{-2}$ s$^{-1}$, in the 0.1 to 100 GeV energy range. 

The spectrum of the $\gamma$-ray emission is shown in Figure \ref{fig:msh1739} (\rig\ panel), where the best-fit power law model to the data is also presented. For energies below 400 MeV above 6.4 GeV only flux upper limits are determined from the data. The best-fit model to the photon distribution yields a spectral index of $\Gamma = Ð2.5 \pm 0.3$, and an integrated flux between 100 MeV and 100 GeV of $F_{\gamma}\approx 2.1\times10^{-7}(9.6\times10^{-11})$ photons/cm$^{2}$/s (erg/cm$^{2}$/s). At the distance estimated from the maser velocity, 12 kpc, the luminosity in this bandpass is hence estimated to be $L_{\gamma}\approx2\times10^{36}$ erg/s.

\section{Discussion}

\noindent The shocks of SNRs are known to accelerate particles to extremely high energies, as evidenced by non-thermal X-ray emission and $\gamma$-ray observations. The population of accelerated particles produce $\gamma$-rays through leptonic mechanisms, inverse-Compton (IC) scattering of ambient photons by energetic electrons, and non-thermal bremsstrahlung, as well as through the decay of neutral pions produced in the collisions between highly energetic hadrons and ambient protons. 

\subsection{Modeling the broadband emission from SNRs}

\noindent To reproduce the observed broadband spectral characteristics of the non-thermal emission from \snrs, we have used models to simulate emission from both electrons and protons accelerated at the shock of each of these SNRs. The spectral momentum distribution $dN_{i}/dp$ of the accelerated particles is adopted to be:
\begin{equation}
\frac{dN_{i}}{dp} = a_{i} \,p^{-\alpha_{i}} \exp\left(-\frac{p}{p_{0\,i}}\right),
\label{eq:uno}
\end{equation}
where $i$ is the particle species (either proton or electron), and $p_{0\,i}$ is the exponential particle momentum cutoff. We then transform the particle distribution to energy space, as well as the exponential cutoff to be defined by an input energy, $E_{0\,i}$. The sum of the integrals of the distributions is set to equal the total energy in accelerated particles within the SNR shell, $E_{\text{CR}}=\theta E_{\text{SN}}$, where $\theta$ is the efficiency of the shock in depositing energy into cosmic rays. The electron to proton ratio at 10 GeV, which is set as an input to the model, then allows us to calculate the values of the coefficients for the proton and electron distributions, $a_{p}$ and $a_{e}$ respectively. The indices of electrons and protons are set to the same value in each model, since analytic and semi-analytic studies suggest this is the case \citep{reynolds_2008}. A simple geometry is assumed for the model, where the SNR is a spherical shell of shock-compressed material, with thickness $R/12$ and compression ratio 4, expanding into a uniform medium of density $n_0$. 

\begin{figure}
\begin{center}
\includegraphics[width=\columnwidth]{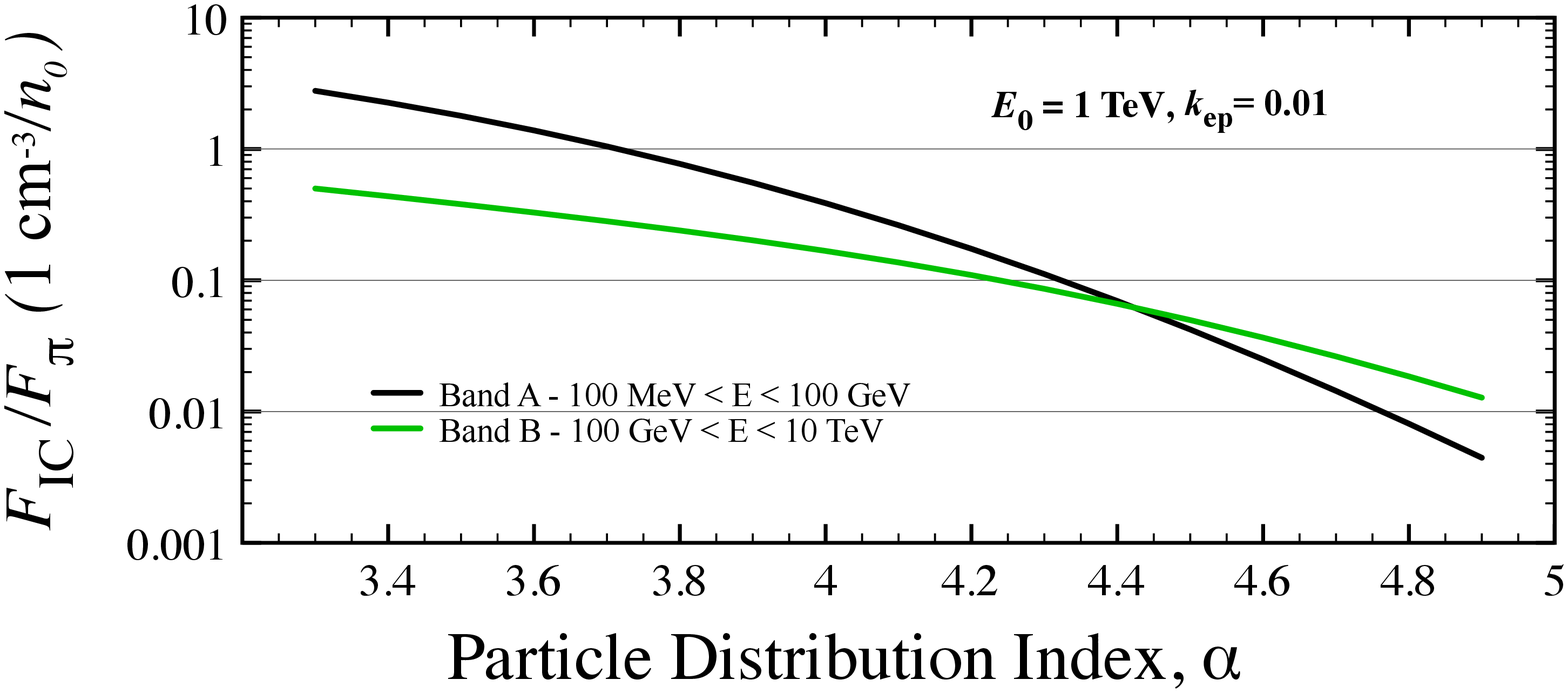}\\
\caption{\footnotesize  Inverse Compton to $\pi^0$-decay flux ratio as a function of the particle momentum distribution index, $\alpha$. The black curve is for the photon energy band 100 MeV to 100 GeV, and the green line indicates the flux ratio in the 100 GeV to 10 TeV range. The exponential particle energy cutoff, for both electrons and protons, has been fixed at 1 TeV, and the electron to proton ratio at 10 GeV is $k_{ep}=0.01$. The ratio is scaled by an ambient density $n_0=1 \text{cm}^{-3}$, since the $\pi^0$-decay component is proportional to this density and the IC flux is not.}
\label{fig:icp-ind}
\end{center}
\end{figure}

\begin{figure}
\begin{center}
\includegraphics[width=\columnwidth]{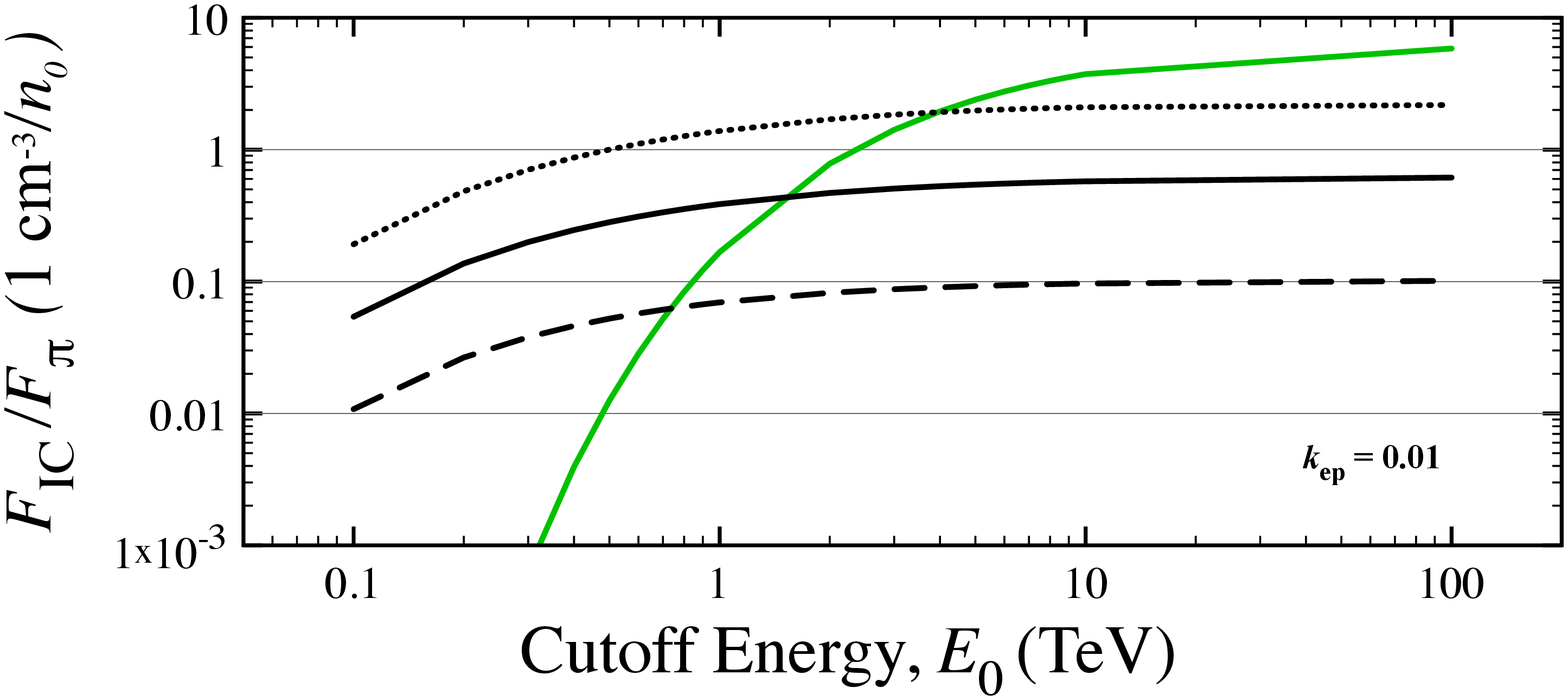}\\
\caption{\footnotesize  Inverse Compton to $\pi^0$-decay flux ratio as a function of the particle distribution exponential energy cutoff. The black curve is for the photon energy band 100 MeV to 100 GeV, and the green line indicates the flux ratio in the 100 GeV to 10 TeV range. The particle distribution spectral energy distribution index $\alpha$, for both electrons and protons, is set to 3.6 ({\it dotted}), 4.0 ({\it solid}) and 4.4 ({\it dashed}). The electron to proton ratio at 10 GeV is fixed at $k_{ep}=0.01$. The ratio is scaled by an ambient density $n_0=1\,\text{cm}^{-3}$, since the $\pi^0$-decay component is proportional to this density and the IC flux is not.}
\label{fig:icp-ecut}
\end{center}
\end{figure}

The $\pi^0$-decay emission mechanism model is based on that from \citet{kamae_2006} using a scaling factor of 1.85 for helium and heavier nuclei \citep{mori_2009}, as described in \citet{castro_snrmc}. The synchrotron and inverse Compton (IC) emission components follow the models presented in \citet[][and references therein]{baring_1999}, and the non-thermal bremsstrahlung emission is modeled using the prescription presented in \citet{bykov_2000}. The ambient photon field used in the inverse Compton model is that of the Cosmic Microwave Background (CMB), $kT_{\text{CMB}}=2.725$~K, and distances to the SNRs established through other observations are adopted. The electron to proton ratio, postshock magnetic field strength ($B_2$), ambient density, and cut-off energy of the particles were adjusted to build the leptonic or hadronic scenarios,
and for the broadband spectrum to fit the observations. While the emission volumes of all the gamma-ray radiation mechanisms is expected to be the approximately the same, if the magnetic field occupies a fraction $f_B<1$ of the volume of the shell, the synchrotron radiation emitting volume will be smaller than those of the others \citep{lazendic_2004}. Since we do not incorporate this filling factor in this single-zone model, the values derived from the fits will underestimate the magnetic field by a factor of $f_B^{-2/(\alpha-1)}$.

In order to constrain the parameter space it is useful to consider which processes might dominate given different particle distribution characteristics. Below we illustrate how the inverse Compton and non-thermal Bremsstrahlung flux compare to the $\pi^0$-decay emission over some of the parameter space.  The relativistic electron to relativistic proton ratio determined in observations of cosmic ray abundances at Earth is $k_{ep}=0.01$ \citep{hillas_2005}. Hence we assume this value when comparing $\pi^0$-decay emission with both the leptonic mechanisms.

Figures \ref{fig:icp-ind} and \ref{fig:icp-ecut} show how the inverse Compton to $\pi^0$-decay flux ratio varies as a function of spectral index and exponential particle energy cutoff respectively. The ratio is scaled by a factor of $1 \text{ cm}^{-3}/ n_0$ since the hadronic component is proportional to the density of the medium, unlike IC emission. Curves are shown for the flux ratios in two bands: black lines denote predictions for energies typical of \fermi\ observations of SNRs, 100 MeV to 100 GeV, and green lines trace those for the energy range typically covered by ground based Cerenkov telescopes, i.e. 100 GeV to 10 TeV.

\begin{figure}
\begin{center}
\includegraphics[width=\columnwidth]{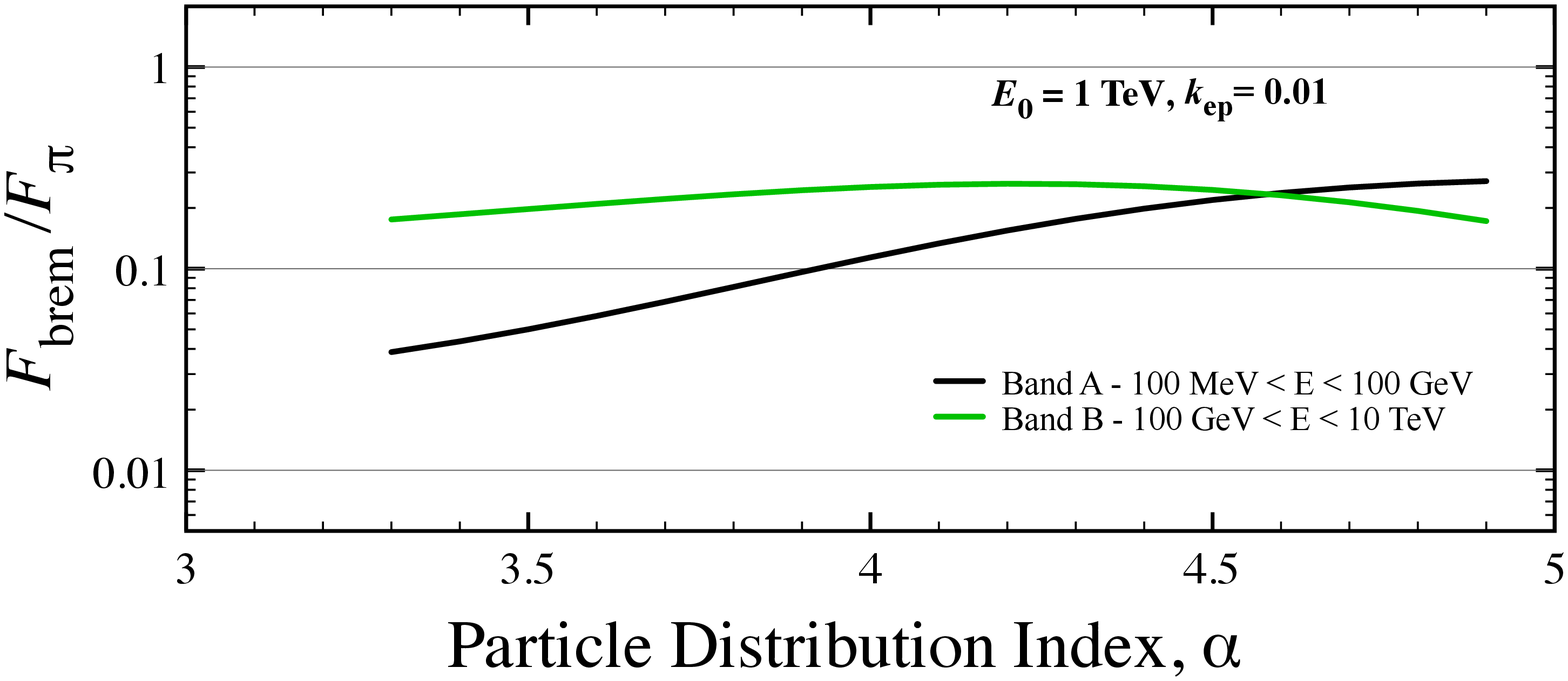}\\
\caption{\footnotesize  Nonthermal Bremsstrahlung to $\pi^0$-decay flux ratio as a function of the particle momentum distribution index, $\alpha$. The black curve is for the photon energy band 100 MeV to 100 GeV, and the green line indicates the flux ratio in the 100 GeV to 10 TeV range. The exponential particle energy cutoff, for both electrons and protons, has been fixed at 1 TeV, and the electron to proton ratio at 10 GeV is $k_{ep}=0.01$.}
\label{fig:bp-ind}
\end{center}
\end{figure}

\begin{figure}
\begin{center}
\includegraphics[width=\columnwidth]{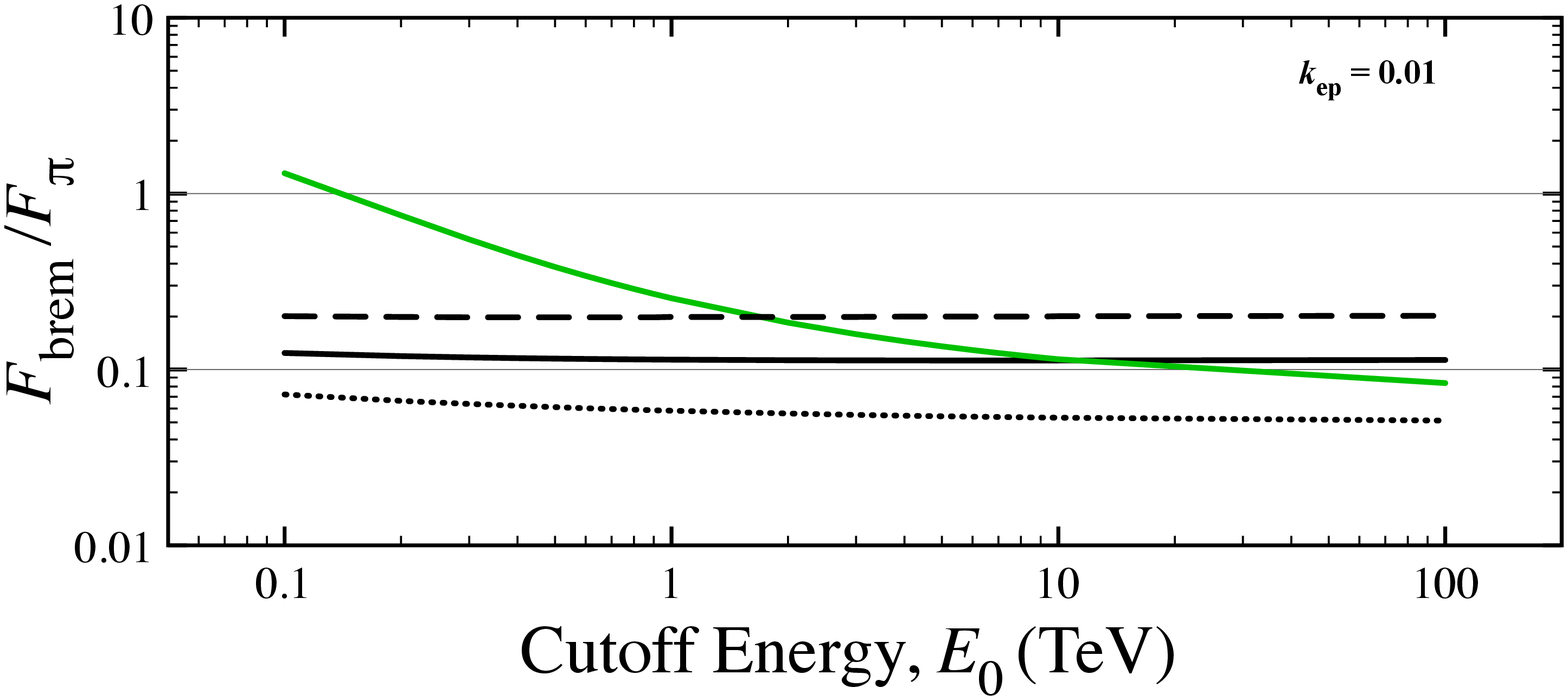}\\
\caption{\footnotesize  Nonthermal Bremsstrahlung to $\pi^0$-decay flux ratio as a function of the particle distribution exponential energy cutoff. The black curve is for the photon energy band 100 MeV to 100 GeV, and the green line indicates the flux ratio in the 100 GeV to 10 TeV range. The particle spectral momentum distribution index $\alpha$, for both electrons and protons, is set to 3.6 ({\it dotted}), 4.0 ({\it solid}) and 4.4 ({\it dashed}). The electron to proton ratio at 10 GeV is fixed at $k_{ep}=0.01$.}
\label{fig:bp-ecut}
\end{center}
\end{figure}

This analysis suggests that if $k_{ep}=0.01$, as expected, and the density of the interstellar or circumstellar medium is on average larger than 1 cm$^{-3}$, the hadronic component is expected to dominate over IC in both the 0.1-100 GeV and the 0.1-10 TeV bands. Only for spectral indices harder than 1.8 and cutoff energies higher than 0.5 TeV does the model predict IC to match or surpass hadronic emission in the LAT energy band. At higher densities, typical in environments such as those of SNRs interacting with molecular clouds, the model suggests emission from pion decay  will be the more significant component of the $\gamma$-ray flux.

The comparison between nonthermal Bremsstrahlung and $\pi^0$-decay flux is shown in Figures \ref{fig:bp-ind} and \ref{fig:bp-ecut}. Since the predicted emission form both mechanisms is proportional to the ambient medium density no scaling is required.

The model predicts that in the \fermi\ energy range, the flux from the accelerated hadrons colliding with the ambient material will dominate over the non thermal Bremsstrahlung for all particle distributions considered. The electron to proton ratio $k_{ep}$ must be larger than $\sim 0.1$ (10 times the value derived from cosmic ray detection experiments) for this leptonic mechanism to significantly contribute to the $\gamma$-ray flux in the MeV-GeV band in comparison to $\pi^0$-emission.

Since we do not expect bremsstrahlung to dominate in the \fermi\ energy band for reasonable sets of parameters, we consider two main scenarios for the origin of the observed $\gamma$-ray flux from each of the SNRs \snrs: sets where the IC emission dominates, and others where $\pi^0$-decay is the main emission mechanism. 

For hadronic models, since the $\gamma$-ray flux is proportional to the ambient density and total energy in cosmic ray protons, these two parameters are degenerate. Hence, we fix the total energy in accelerated particles at $E_{\text{CR}}\equiv E_{\text{CR,p}}+E_{\text{CR,e}}=4\times10^{50}$ erg. This value represents a reasonable upper limit for the fraction of energy deposited in CRs for a supernova remnant produced by a $10^{51}$ erg explosion. For the inverse Compton dominated cases we vary the total CR
energy and $k_{ep}$ so as to simultaneously fit the data and limit the hadronic emission contribution to a maximum of 20\% of the total $\gamma$-ray flux in the \fermi\ range. 

The IC $\gamma$-ray flux might be enhanced by the presence of additional photon fields to the CMB, such as the infrared (IR) emission from interstellar dust ($T\sim5$ K) or ambient starlight \citep{strong_2000}, or IR emission from warm dust in the SNR itself. For each of the \fermi\ sources studied, we briefly comment on how adding these additional components impacts the results of the model.

\subsection{SNR \wf\ and its field}
\subsubsection{Emission from the SNR}

\begin{table*}
\begin{center}
\begin{threeparttable} 
\caption{Input Model Parameters}
\label{tab:had}
\begin{tabular}{cccccccccccc}
\toprule
\noalign{\smallskip}
\multirow{2}{*}{SNR}&\multirow{2}{*}{Model}&\multirow{2}{*}{$\alpha$}&$E_{\text{0,e}}$&$E_{\text{0,p}}$&\multirow{2}{*}{$k_{ep}$}&$n_0$&$B_2$\tablenotemark{a}&$d$&$E_{\text{CR,p}}$&$E_{\text{CR,e}}$\\
&&&\multicolumn{2}{c}{(TeV)}&&(cm$^{-3}$)&($\mu$G)&(kpc)&\multicolumn{2}{c}{($10^{50}$ erg)}\\\noalign{\smallskip}
\midrule
\multirow{2}{*}{\wf} & {\it Hadronic} & 4.00 & 13 & 13 & 0.001 & 4.00 & 180 & 4.2 & 3.994 & 0.006 \\ 
& {\it Leptonic} & 3.95 & 4.5 & 4.5 & 0.13 & 9.00 & 35.0 & 4.2 & 0.365 & 0.065  \\
\midrule
\multirow{2}{*}{\ctbt} &{\it Hadronic} & 4.00 & 0.1 & 0.1 & 0.01 & 60.0 & 9.0 & 11  & 3.912 & 0.088 \\
& {\it Leptonic} &3.60 & 0.3 & 0.3 & 0.5 & 0.08 & 0.2 & 11  & 17.87 & 9.128 \\

\midrule
\multirow{2}{*}{\msh} & {\it Hadronic} & 4.00 & 0.1 & 0.1 & 0.01 & 18.0 & 140 & 12 & 3.912 & 0.088 \\
&{\it Leptonic} & 3.95 & 0.3 & 0.3 & $0.5$ & 0.50 & 3.0 & 12 & 9.722 & 8.278 \\

\bottomrule
\noalign{\smallskip}
\end{tabular}
\begin{tablenotes}[para]
\item[a]{$B_2$ is the magnetic field immediately behind the shock, and is scaled by a factor $f_B^{-2/(\alpha-1)}$, to account for the magnetic field occupying a fraction $f_B$ of the SNR shell volume.}

\end{tablenotes}
\end{threeparttable} 
\end{center}
\end{table*}

\begin{figure}
\begin{center}
\includegraphics[width=\columnwidth]{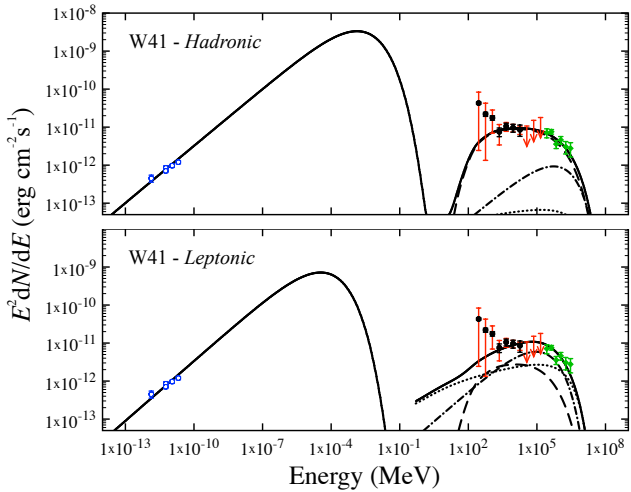}\\
\caption{\footnotesize  Broadband fits to radio \citep[blue squares,][]{tian_2007}, \fermi\ (black circles), and HESS \citep[green diamonds,][]{aharonian_2006} observations of \wf\ with the hadronic ({\it top}), and leptonic ({\it bottom})) models. The solid black line is the total nonthermal emission predicted by the model. The modeled spectra from inverse Compton emission (dot-dashed), $\pi^0$-decay (dashed), and nonthermal bremsstrahlung (dotted), are also shown.}
\label{fig:w41-spec}
\end{center}
\end{figure}

\noindent In order to determine the origin of the $\gamma$-ray emission in the direction of \wf\ we fit its broadband spectrum using the model described above. Figure \ref{fig:w41-spec} shows the model fits to the broadband emission observed from SNR \wf. The radio spectrum is a combination of observations at 20 cm, 10 cm and 6 cm \citep{altenhoff_1970}, 90cm \citep{kassim_1992}, and additional data at 20cm \citep{tian_2007}. The HESS very high energy $\gamma$-ray spectrum is that obtained by \citet{aharonian_2006}. The solid black line is the total nonthermal emission predicted by the model. The modeled spectra from IC emission (dot-dashed), $\pi^0$-decay (dashed), and nonthermal bremsstrahlung (dotted), are also shown.

The two different sets of parameters required for the leptonic and hadronic scenarios are summarized in Table \ref{tab:had}. The hadronic scenario is the most favorable since the fit is adequate and all parameters are within reasonable ranges. As described in $\S$3.1, the mean ambient density derived is $n_0\sim4\, (4\times10^{50}\text{ erg}/E_{CR}) (d/4.2\text{ kpc}) \text{ cm}^{-3}$. Given the distance and the SNR radius determined through radio observations $R_{\text{W41}}\approx18 (d/4.2\text{ kpc})$ pc \citep{leahy_2008}, this mean density would imply the shock has swept-up $M_{sw}\gtrsim3\times 10^3 \text{M}_\odot$. If we then assume the supernova explosion energy available for shock heating and expansion to be 0.6$\times10^{51}$ erg, and we use the simple analytic treatment prescribed in \citet{castro_2011}, the resulting age and shock velocity are $t\sim7\times10^4$ years and $V_S\sim110$ km/s, respectively. Such a low shock velocity would not be expected to result in significant thermal X-ray emission, where, even at thermal equilibrium among shock heated particles, the resulting electron temperature would be $kT_e\sim10^{-2}$ keV.

This parameter set requires the exponential cutoff energy of the proton distribution to be $\sim13$ TeV. However, the cutoff for the electron particle distribution is not constrained by the observational data obtained up to date (the results quoted in Table \ref{tab:had} and illustrated in the top panel of Figure \ref{fig:w41-spec} are for $E_{\text{0,e}}=E_{\text{0,p}}$). 

The fit for the leptonic dominated model requires a total energy in cosmic ray electrons $E_{\text{CR,e}}=6.5\times10^{48}$ erg. The spectral shape of the \fermi\ and HESS data points requires contributions from both the inverse Compton and nonthermal Bremsstrahlung for an appropriate fit. Therefore, since the density and cosmic ray electron energy are constrained by the data, for the hadronic component not to disrupt the fit to the observed $\gamma$-rays, the electron to proton ratio required is an order of magnitude larger than expected, i.e. $k_{ep}\gtrsim 0.13$. 

To investigate the effects of additional photon fields to the CMB on the IC dominated scenario, the observed $\gamma$-ray flux was fit by setting the electron to proton ratio at $k_{ep}=0.01$ and the total energy in cosmic rays at $E_{cr}=4\times10^{50}$ erg, and all other parameters at the values shown for the leptonic model in Table \ref{tab:had}. The required photon field energy density obtained was 20 times that of the CMB, for a  component with temperature $T=25$K (characteristic of interstellar dust). This is higher than expected from such photon field anywhere in our Galaxy \citep{strong_2000}. Both ambient starlight and emission from warm SNR dust have higher characteristic temperatures, and hence even larger photon field energy densities would be required. We conclude that from the two scenarios where the origin of the $\gamma$-ray emission detected by HESS and \fermi\ results from particles accelerated at the shock of \wf, the hadronic case is certainly the most compelling.

\subsubsection{Other possible $\gamma$-sources in the region}

\noindent There are several other sources in the field which should be considered as possibly associated with the $\gamma$-ray emission detected with \fermi\ and HESS. Below we outline these alternative hypotheses and evaluate their merits.

The 85 ms pulsar PSR J1833--0827 lies to the NW of SNR W41, approximately 20$'$ away from the centroid of the \fermi\ source. \citet{gaensler_1995} concluded that it was possibly associated with the supernova remnant and its distance, 4-5 kpc derived kinematically from H {\footnotesize I} absorption observations \citep{weisberg_1995}, is consistent with such scenario. However, the spin down luminosity of this pulsar is $\dot{E}=5.8\times10^{35}$ erg/s, approximately all of which would be required to power the $\gamma$-ray emission detected with \fermi, $L_{\gamma}\approx5\times10^{35}$ erg/s. Moreover, the positions of the HESS source and the \fermi\ source do not seem consistent with being associated with this pulsar.

There is a compact X-ray source XMMU J183435.3--084443/CXOU 183434.9--084443 located at the center of W41 and within the extent of the HESS and \fermi\ emission \citep{mukherjee_2009,misanovic_2011}. The X-ray emission of this source appears to have a strongly absorbed power law spectrum, and \citet{misanovic_2011} detected evidence of compact, $r\lesssim20''$ extended emission resolved around a point-like source. \citep{mukherjee_2009} proposed that this source could be a pulsar surrounded by a PWN, and that it would be responsible for the TeV $\gamma$-ray emission detected. In the PWN scenario, the energetics of the system can be the estimated combining the observed X-ray luminosity, $L_X\sim4\times10^{33}$ erg s$^{-1}$ at an estimated distance of 4 kpc, and the $L_X$--$\dot E$ relation \citep[e.g.][]{kargaltsev_2008}. Hence, the spin-down power of the putative pulsar is estimated to be $\dot E\sim10^{36}-10^{37}$ erg s$^{-1}$. The luminosity in the \fermi\ range, $L_{\gamma}=2\times10^{35}$ erg s$^{-1}$, would require an efficiency of $L_{\gamma}/\dot E\sim 10\%$. While it is not possible to rule this scenario, it appears an unlikely association to the $\gamma$-ray source given large relative extent of these compared to the X-ray emission.

Most recently, an additional high energy source has been detected in this already crowded region, the magnetar \magj. A short burst from this source triggered the {\it Swift} Burst Alert Telescope (BAT) in August 2011 \citep{delia_2011}. Approximately 3.3 hr later another burst resembling a soft gamma-ray repeater (SGR) triggered the {\it Fermi} Gamma Ray Burst Monitor (GBM) \citep{guiriec_2011}. Follow up observations have determined pulsation period from this source to be $P\sim2.48$ s \citep{gogus_2011}, its period derivative $\dot{P}=8\times10^{-12}$ s s$^{-1}$ and magnetic field $B=1.4\times10^{14}$ G \citep{kargaltsev_2012}. \citet{younes_2012} detected extended, $r\sim2.5'$ X-ray emission around the magnetar using \chandra, and they argued that this could be understood as evidence of a magnetar wind nebula around \magj. The scenario where the $\gamma$-ray emission detected in this region results from either the magnetar or its wind nebulae is unlikely because, similarly to XMMU J183435.3--084443/CXOU 183434.9--084443, the extension of this X-ray source does not seem to match that detected in the GeV range.

Additionally, we performed a pulsation search for the magnetar in $\gamma$-rays
using \fermi\ data from the start of mission to November 30, 2012.
We extracted events from within a search radius of 2 degrees centered
on the magnetar, using an energy range of 100~MeV to 300~GeV and a
zenith angle maximum of 100 degrees.  Using the ephemeris from
\citet{kargaltsev_2012}, we searched for pulsations using {\it gtpsearch}.
Based on the uncertainties in $P$ and $\dot{P}$, we searched a
period range from 2.4815439~s to 2.4815465~s (at the start of the
time series), corresponding to $\sim 58$ Fourier-independent periods,
oversampled by a factor of $\sim 3$. For the nominal period, the
full uncertainty range in $\dot{P}$ corresponds to a shift of nearly
0.75 in phase over the full length of the observation. We thus
performed the search for the full $\dot{P}$ uncertainty range using
6 equally-spaced steps in $\dot{P}$.

Using a $Z_N$ test for a single harmonic, the search yielded a
maximum statistic of $Z_2= 10.87$ at a period of 2.4815444~s with
a chance probability of $\sim 0.77$ after accounting for trials in
the $P$ and $\dot{P}$ search.  We therefore find no statistically
significant evidence for $\gamma$-ray pulsations from the magnetar.
We cannot rule out steady emission from the magnetar, but consider
it more likely that the $\gamma$-rays originate from \wf.

\subsection{SNRs \ctbt\ and \msh}

\begin{figure}
\begin{center}
\includegraphics[width=\columnwidth]{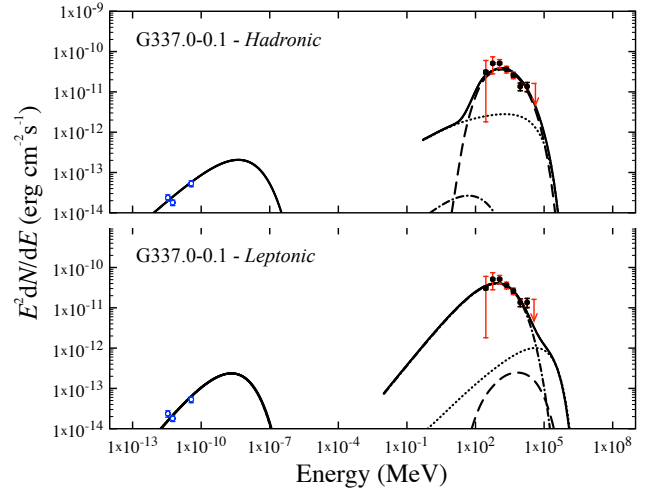}
\caption{\footnotesize  Broadband fits to radio \citep[blue squares,][]{tian_2007}, and \fermi\ (black circles) observations of \ctbt\ with the hadronic ({\it top}), and leptonic ({\it bottom})) models. The solid black line is the total nonthermal emission predicted by the model. The modeled spectra from inverse Compton emission (dot-dashed), $\pi^0$-decay (dashed), and nonthermal bremsstrahlung (dotted), are also shown.}
\label{fig:ctb33-spec}
\end{center}
\end{figure}

\begin{figure}
\begin{center}
\includegraphics[width=\columnwidth]{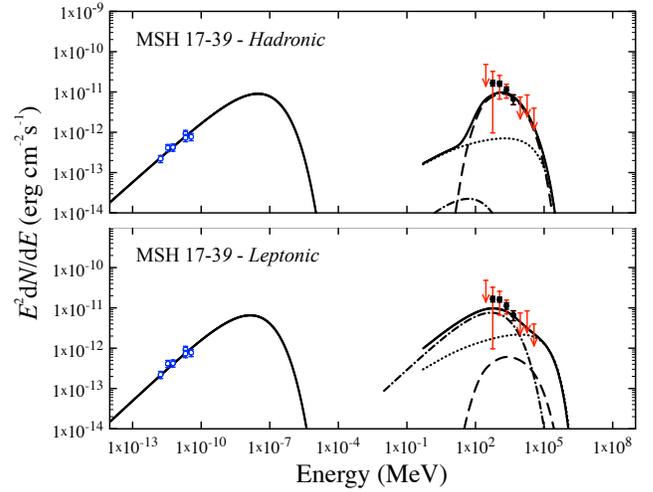}
\caption{\footnotesize  Broadband fits to radio \citep[blue squares,][]{tian_2007}, and \fermi\ (black circles) observations of \msh\ with the hadronic ({\it top}), and leptonic ({\it bottom})) models. The solid black line is the total nonthermal emission predicted by the model. The modeled spectra from inverse Compton emission (dot-dashed), $\pi^0$-decay (dashed), and nonthermal bremsstrahlung (dotted), are also shown.}
\label{fig:msh17-39-spec}
\end{center}
\end{figure}

\noindent The broadband fits to the observations of \ctbt\ and \msh\ are shown in Figures \ref{fig:ctb33-spec} and \ref{fig:msh17-39-spec} respectively, and the best fit parameters are presented in Table \ref{tab:had}. The radio data of \ctbt\ are those obtained by \citet{sarma_1997} from ATCA observations, and the radio spectrum of \msh\ is a combination of observations with the MOST, Mills Cross and Parkes 64m telescopes \citep{dickel_1973,caswell_1975,milne_1975,caswell_1980,gray_1994}.

In both \ctbt\ and \msh\ the hadronic scenarios are successful in fitting the observations and provide reasonable parameter sets. The ambient densities derived assuming the observed $\gamma$-ray emission results from $\pi^0$-decay are approximately $60d_{11}$ and $18d_{12}$ cm$^{-3}$ respectively, given a total energy in cosmic rays of $4\times10^{50}\text{ erg}$. 

In contrast to the hadronic model fits, inverse Compton dominated scenarios, for both \ctbt\ and \msh, require total energies in accelerated electrons ($E_{\text{CR,e}}\sim10^{51}$erg) which are much higher than expected. Hence leptonic emission is not likely to be the dominant mechanism behind the observed $\gamma$-rays. Even if we assume an electron to proton ratio $\sim0.5$, approximately 50 times the ratio derived from CR observations at Earth, the resulting total energy in cosmic rays would far exceed the canonical value of ejecta kinetic energies of supernova explosions, also $10^{51}$ erg. There are uncertainties related to distances derived using Galactic rotational curves and kinematic information, especially for objects at longitudes close to the Galactic center, and hence some of the derived parameters could be somewhat different. For example, if the distances to these SNRs are smaller by a factor of 3, the total energy in cosmic ray electrons required to fit the observations would be $\sim10^{50}$ erg. Considering the electron to proton ratio observed at Earth such electron energies would still suggest unreasonably large amounts of total energy in cosmic rays.

In order to investigate the effect of additional photon fields to the CMB on IC dominated cases, the observed $\gamma$-ray flux of the remnants was fit by setting the electron to proton ratio at $k_{ep}=0.01$ and the total energy in cosmic rays at $E_{cr}=4\times10^{50}$ erg, and all other parameters at the values shown for leptonic models in Table \ref{tab:had}. The required photon field energy density obtained was 600 and 2000 times the CMB energy density for \ctbt\ and \msh\ respectively, for a  component with temperature $T=25$K (characteristic of interstellar dust). These are much higher than expected from such photon fields anywhere in our Galaxy. Both ambient starlight and emission from warm SNR dust have higher characteristic temperatures, and hence even larger photon field energy densities would be required.

\section{Summary}

Using observations with the {\it Fermi}-LAT, we have reported and studied $\gamma$-ray emission from three SNRs known to be interacting with molecular clouds, \snrs. 
Their radio and $\gamma$-ray spectra are fit using a broadband model of non-thermal emission from SNRs. This is a simple, spherically symmetric model which allows for approximate conclusions to be drawn about the origin of the observed emission. In all three cases a hadronic origin for the $\gamma$-ray flux observed is preferred by the model to leptonic scenarios. In the case of \wf, other possible sources of the MeV-GeV emission are considered, such as a magnetar and a pulsar wind nebulae in the field, and while these cannot be ruled out completely, the nature of the \fermi\ observations favors the SNR origin.

\acknowledgments
The authors thank Stefan Funk, Elizabeth Hays, Joshua Lande for their advise regarding {\it Fermi}-LAT data analysis. We also thank Dale Frail for conversations about SNR-MC interaction. Additionally, The authors thank the referee for constructive comments and recommendations which have strengthened the paper. This work was partially funded by NASA Fermi grant NNX10AP70G. 

DC also acknowledges support for this work provided by the National Aeronautics and Space Administration through the Smithsonian Astrophysical Observatory contract SV3-73016 to MIT for Support of the Chandra X-Ray Center, which is operated by the Smithsonian Astrophysical Observatory for and on behalf of the National Aeronautics Space Administration under contract NAS8-03060. POS acknowledges partial support from NASA contract NAS8-03060.

\vspace{1cm}
\bibliography{ref}

\end{document}